\documentclass[review]{elsarticle}

\usepackage{lineno,hyperref}
\modulolinenumbers[5]

\journal{Journal of \LaTeX\ Templates}

\begin{document}

\begin{frontmatter}

\title{Bayesian Nonparametric Model for Weighted Data Using Mixture of Burr XII Distributions}

\author{S. Bohlourihajjar}

\address{bohlurihajjar.soghra@razi.ac.ir}
\author{ S. Khazaei}
\address{s.khazaei@razi.ac.ir}

\begin{abstract}
Dirichlet process mixture model (DPMM) is a popular Bayesian nonparametric model. In this paper, we apply this model to weighted data and then estimate the un-weighted distribution from the corresponding weighted distribution using the metropolis-Hastings algorithm. We then apply the DPMM with different kernels to simulated and real data sets. In particular, we work with lifetime data in the presence of censored data and then calculate estimated density and survival values.
\end{abstract}


\end{frontmatter}


\section{Introduction}
Let $X$ be a non-negative random variable with density function $f(x)$ and $w(x)$ be a non-negative function of $x$. A new random variable $X^w$ with density function $g(x)$ that is defined as bellow,
\begin{equation}\label{model}
g(x)=\frac{w(x)f(x)}{E[w(X)]},~~~E[w(X)]<\infty,~~~x \geq 0,
\end{equation}
\newline
is called weighted random variable with respect to X, and $g(x)$ is called the weighted density function  with respect to $f(x)$. Especially, if $w(x)=x$, the resulted weighted distribution is called length-biased distribution which has various applications in many areas. Zelen and Feinleib(1969) applied the length-biased distribution to detect breast cancer early, and also Patil et al. (1977) used this weighted distribution to study human families and wild-life population.
Patil et al.(1978) introduced distributions of the type given in equation \ref{model} with an arbitrary non-negative weight function $w(x)$ and gave practical examples. For more examples of weighted distributions and their applications, see \cite{blumenthal63},\cite{patil77},\cite{patil78},\cite{mahfoud82},\cite{Gupta90}.

In this work we consider wighted distributions nonparametric Bayesian methods by using Dirichlet process mixture model(DPMM) which is a papular Bayesian nonparametric model and then, appply these idea to survival analysis.

In \cite{Hatjispyros}, DPMM is used for density estimation under length-biased data. They consider a log-normal distribution as the kernel function which its shape parameter has the determined distribution.
Here we consider a DPMM with Burr(XII) distribution as the kernel function with two parameters (scale parameters) whose distributions are random in the model.
Since the support of the Burr(XII) distribution is $R^+$, it is suitable for survival study (\cite{Bohlourihajjar},\cite{lanjoni},\cite{rao15},\cite{rodriguez}). In \cite{Bohlourihajjar} Burr(XII) distribution is used  as the kernel in DPMM, and then, survival function and hazard rate are calculated for simulated and real data.

In the next section, preliminary concepts and the methodology are presented, and section 3 describes the nonparametric Bayesian approach that allows us to define the model which will be used in the next section.
Section 4 is contained the modeling and algorithm that is used for the sampling. In the next section, we consider data illustrations for applying the model to the data. Finally, we summarized our results in the conclusion section.

\section{Preliminary and Methodology}
We want to estimate the density and other survival functions by considering a general case of weight function $w(x)$. The strategy for avoiding computing the intractable normalizing constant would be to model $g(x)$ directly and then make inference about $f(x)$ by considering this fact that $g(x)\propto w(x)f(x)$.
If we set $f(x;\theta)$ as a parametric family, so $f(x;\theta)$ and $g(x;\theta)$ are known except the normalizing constant that may be not tractable.

Let $w(.)$ be a general weight function, an essential condition to model F(.) through G(.) \bigg{(}F(.) and G(.) denote the distribution functions of f(.) and g(.) respectively.\bigg{)} is

\begin{equation}\label{condition}
\int_0^\infty w(x)^{-1} g(x) dx<\infty,
\end{equation}
\newline because f is a distribution function.

 Through invertibility, equation (\ref{condition}) enables us to reconstruct F from G.

In the Bayesian nonparametric framework, we assign an appropriate nonparametric prior distribution on g, providing relation (\ref{condition}). The question that now arises is how the posterior structures obtained after modeling g directly can convert to the posterior structures from f.

 The first step is to construct a method to convert a weighted sample to an un-weighted one. Then it is possible to inference about the posteriors.

An indirect method to simulate samples of complex distribution is Monte Carlo Markov chain (MCMC) approach. Metropolise-Hastings algorithm \cite{metro95} is one of the MCMC methods which simulates samples from a probability distribution by making use of the full joint density function and proposal distributions for each of the variables of interest.

In general form, the Metropolis-Hastings algorithm is as the following form,
\begin{eqnarray*}
&&\hspace{-5cm}Algorithm 1: ~~Metropolis-Hastings~algorithm\\
&&\hspace{-4.5cm} Initialize with~x^{(0)}\sim q(x)\\
&&\hspace{-4cm}\textbf{for}~i=1,2,... ~\textbf{do}\\
&&\hspace{-4cm}~Propose~x^{cand}\sim q(x^{(i)}|x^{(i-1)})\\
&&\hspace{-4cm}~Calculate~the~acceptance~probability:  \\
&&\hspace{-4cm}~~~~~\alpha(x^{cand}|x^{(i-1)})=min\{1,\frac{q(x^{(i-1)}|x^{cand})\pi(x^{cand})}{q(x^{cand}|x^{(i-1)})\pi(x^{(i-1)})}\}\\
&&\hspace{-4cm}~Generate~u\sim Uniform(u;0,1)\\
&&\hspace{-4cm}~~\textbf{if}~u<\alpha ~\textbf{then}\\
&&\hspace{-4cm}~~~ ~x^{(i)}\leftarrow ~x^{cand}\\
&&\hspace{-4cm}~~\textbf{else}\\
&&\hspace{-4cm}~~~~x^{(i)}\leftarrow x^{(i-1)}\\
&&\hspace{-4cm}~~\textbf{end if}\\
&&\hspace{-4cm}\textbf{end for}
\end{eqnarray*}

Hatjispyros and et al. used the Metropolis-Hastings algorithm to convert a length biased sample to the unbiased version.
Now, we want to apply this algorithm by using general weight function satisfied in relation (\ref{condition}) to convert the sample from weighted distribution to the un-weighted version.

Suppose that $y_1,y_2,...,y_N$ denote a random sample of $g$. The Metropolis-Hastings algorithm converts this sample to a sample from $f(x)\propto w(x)^{-1}g(x)$.
We assume $g(.)$ is replaced by $q(.)$ in the algorithm 1 with acceptance probability $min \{1,\frac{w^{-1}(y_{j+1})}{w^{-1}(x_j)}\}$.
If $x_j$ denotes the current sample from f(x), then
\begin{eqnarray}\label{rejection}
&&x_{j+1}=y_{j+1} ~~~~with ~probability ~~~~min \{1,\frac{w^{-1}(y_{j+1})}{w^{-1}(x_j)}\},\\
&&x_{j+1}=x_j ~~~~otherwise.\nonumber
\end{eqnarray}

The transition density is,
$$p(x_{j+1}|x_j)=min\{1,\frac{w^{-1}(y_{j+1})}{w^{-1}(x_j)}\}g(x_{j+1})+\{1-r(x_j)\}1(x_{j+1}=x_j),$$
where
$$r(x)=\int min \{1,\frac{w^{-1}(x^*)}{w^{-1}(x)}\}g(x^*)dx^*.$$
We can have the following outline methodology in a general form:

1- $(y_1,...,y_n)$ is a sample from g that we are going to assign a suitable nonparametric prior to it.

2- Using MCMC methods, the posterior values of the random measure $\Pi(dg|y_1,...,y_n)$ and other relevant parameters will obtain. So a sequence $\{y^{l}_{n+1}\},l=1,2,...$, from the posterior predictive density $g(y|y_1,...,y_n)$ will be generated.

3- $\{y^{l}_{n+1}\}$ will form a sequence of proposal values of a Metropolis-Hastings chain with
the stationary density of the weighted posterior predictive i.e. $\{y^{l}_{n+1}\}\propto w(y)^{-1}g(y|y_1,...,y_n)$.
By the equation (\ref{rejection}) we generate the $\{x^{l}_{n+1}\}$ values at the level l.

4- Then $\{x^{l}_{n+1}\}$ values are a sample from the posterior of predictive f (un-weighted density).

\section{The model and inference}
Modeling $g(x)$, the weighted distribution, in the Bayesian nonparametric framework is based on infinite mixture model \cite{Lo84} as the following form

\begin{equation}\label{prior}
g_\mathcal{P}(y)=\int \kappa(y;\theta)P(d\theta),
\end{equation}
\newline where P is a discrete probability measure and $\kappa(y;\theta)$ is a kernel density on $(0,\infty)$ for all $\theta$'s in the parameter space, satisfying the following condition,
$$\int_0^\infty w^{-1}(y)\kappa(y;\theta)dy<\infty.$$
By choosing Burr(XII) density with two parameters c and k as the kernel of the mixture model, so we have
$$g_{c,k,P}(y)=\int_{\mathcal{R}}Burr_{XII}(y|c,k)P(dc,dk),$$
where $P$ is a discrete random probability measure. Suppose $P\sim DP(\upsilon,P_0)$ where $DP(\upsilon,P_0)$ denotes the Dirichlet process with precision parameter $\upsilon >0$ and base measure $P_0$ \cite{fergueson83}. We named this mixture model by Dirichlet process Burr(XII) mixture model (DPBMM).

Hierarchical representation of DPBMM can be presented as the following form,

\begin{eqnarray}
y|c,k&\sim & Burr_{XII}(y|c, k),\nonumber\\
(c,k)|P &\sim & P,\nonumber\\
P|\upsilon, P_0 &\sim & DP(\upsilon,P_0).\\
\nonumber
\end{eqnarray}
Suppose that the base distribution $P_0$ is the prior distribution for the joint distribution of $c$ and $k$.
  By choosing Burr(XII) distribution as the kernel, $P_0$ that yields closed-form expression for $\int k_B(.|c,k) P_0(dc,dk)$ is not available. Moreover, we choose multiple distributions of uniform$(0,\phi)$ and exponential with the parameter $\gamma$ for the $P_0$, i.e.
\begin{equation}
P_0(c,k|\phi,\gamma)=Unif(c|0,\phi)\times Exp(k|\gamma).
\end{equation}
This choice achieves determined goals. By considering hyper-parameters, $\gamma$ and $\phi$ are random, we choose prior distributions $Pareto(a_\phi,b_\phi)$ and $IGamma(a_\gamma,b_\gamma)$ for them respectively.
\newline Set the $a_\phi=a_\gamma=d$ and also d=2 since this value makes the variance of Pareto distribution infinite that cover all values in R. $b_\phi$ and $b_\gamma$ are determined by the data \cite{Bohlourihajjar}.

Finally, for any $t_i, i=1,...,n$, lifetime data in a sample of n observations, by considering DPBMM and selecting priors for parameters of the model we have,
\begin{eqnarray}
t_i|c_i,k_i&\sim & Burr_{XII}(t_i|c_i, k_i),\quad i=1,...,n,\nonumber\\
(c_i,k_i)|\mathcal{P} &\sim & \mathcal{P},\nonumber\\
\mathcal{P} &\sim & DP(\nu,P_0),\nonumber\\
P_0|\gamma,\phi &\sim & Unif(c|0,\phi)\times Exp(k|\gamma)\\
\nu, \gamma, \phi &\sim& Gamma(a_\nu,b_\nu)\times IGamma(a_\gamma,b_\gamma)\times Pareto(a_\phi,b_\phi).\nonumber\\
\nonumber
\end{eqnarray}

 After determining the model, we want to formulate how to sample from DPMMs by Gibbs sampling. According to \cite{kottas6}, Gibbs sampling for drawing a sample from $[(\theta_1,...,\theta_n),\upsilon,...|t]$ based on the following full conditional distributions ( bracket is used to show the conditional and marginal distributions):
\begin{eqnarray}
&& (1)~ [(\theta_i)|(\theta_{-i},z_{-i}),\upsilon,...,t],\quad for ~i=1,...,n\nonumber\\
&& (2)~ [(\theta_j^*)|z,n^*,\upsilon,...,t], ~~for~ j=1,...,n^*\\
&& (3)~ [\upsilon|\{(\theta_j^*),j=1,...,n^*\},n^*,t],[...|\{(\theta_j^*),j=1,...,n^*\},n^*,t].\nonumber
\end{eqnarray}
where t is the vector of failure time data.
Here, $\theta_i$'s are parameters of the kernel in DPMMs that will be analyzed.
\newline
Model (4) and discreteness property of Dirichlet process, exhibit a clustering in $\theta$'s.
We present $n^*$ as the number of the clusters between $\theta_i$'s that denote by $\theta_j^*$'s.
The vector of indicators $z=(z_1,...,z_n)$ indicates the clustering configuration such that, $z_i=j$ when $\theta_i=\theta_j^*$.
Also, the $\theta_{-i}$ that used in (8), will be defined by $\theta_{-i}=(\theta_1,\theta_2,...,\theta_{i-1},\theta_{i+1},...,\theta_n)$.

\section{Modeling}
For modeling the un-weighted density $f(x)$ from the weighted density $g(x)$, we apply the following algorithm.
At first, to generate a sample from $g(x)$, it needs to estimate the parameters of the model. To this aim we draw a sample from $(c_i,k_i)$ and update $z_i$ for each $t_{i}$.

In simulation-based parameter estimation, we use the Gibbs sampler that it includes two steps to reach the goal.
\begin{eqnarray*}
&&\hspace{-7cm}Algorithm 2: Gibbs~sampler\\
&&\hspace{-6cm} 1.~ Initialize~with~~\theta^{(0)}\sim f(\theta)\\
&&\hspace{-6cm} 2.~ For~~i=1,2,...~do\\
&&\hspace{-6cm} ~~~~~~\theta_1^{(i)}  \sim f(\theta_1|\theta_2^{(i-1)},\theta_3^{(i-1)},...,\theta_d^{(i-1)},D),\\
&&\hspace{-6cm} ~~~~~~\vdots  \\
&&\hspace{-6cm} ~~~~~~ \theta_d^{(i)} \sim f(\theta_d|\theta_1^{(i)},\theta_2^{(i)},...,\theta_{d-1}^{(i)},D),
\end{eqnarray*}
where $\theta_1,...,\theta_d$ are model parameters and D is the vector of observations. The values of iteration i would be sampled from the distribution with last version of the other parameter values.

Now, the model will be applied for lifetime data with the presence of right censored data, that is very common in the survival study. To calculate the related distributions we divide data into censored and uncensored observations.

1- Uncensored data:

For uncensored data($t_{io}$), the conditional posterior density of $(c_{i},k_{i})$  is a mixed distribution \cite{neal},
$$f(c_i,k_i|\{(c_{i'},k_{i'});i\neq i'\},\nu, \gamma, \phi,t_{io})=\frac{q^o_0 h^o(c_i,k_i|\phi,\gamma,t_{io})+\sum_{j=1}^{n^{*(i)}}n^{*(i)}_j q^o_j\delta_{c^*_j,k^*_j}}{q_0^o+\sum_{j=1}^{n^{*(i)}}n^{*(i)}_j q^o_j},$$
where $q_j^o=k_B(t_{io}|c_j^*,k_j^*)$ and
\begin{eqnarray*}
q_0^o &=& \nu \int_0^\phi \int_0^\infty k(t_{io}|c,k)G_0(c,k)dcdk\\
&=& \frac{\nu}{\phi} \int_0^\phi \frac{ct_{io}^{c-1}}{(1+t_{io}^c)}(\int_0^\infty\frac{ke^{-\frac{k}{\gamma}}}{(1+t_{io}^c)^k}dk)dc\\
&=& \frac{\nu}{\phi} \int_0^\phi \frac{ct_{io}^{c-1}}{(1+t_{io}^c)(ln(1+t_{io}^c)+\frac{1}{\gamma})}dc
\end{eqnarray*}
in which the last integration can be computed numerically and
$$h^o(c_i,k_i|\gamma,\phi, t_{io})\propto k_B(t_{io}|c_i,k_i)P_0(c_i,k_i|\gamma,\phi)\propto [c_i|\gamma,\phi,t_{io}][k_i|c_i,\gamma,\phi,t_{io}]$$
where
$$[c_i|\gamma,\phi,t_{io}]\propto c_i t_{io}^{c_i-1}I_{(0,\phi)}(c_i)~~~~~i=1,...,n$$
and
$$[k_i|c_i,\gamma,\phi,t_{io}]\propto Gamma(.|2,\frac{1}{[\frac{1}{\gamma}+ln(1+t_{io}^{c_i})]}).$$

2- right censored data:

For right censored data $(t_{ic})$, the conditional posterior density of $(c_i,k_i)$ is
$$f(c_i,k_i|\{(c_i,k_i);i\neq i'\},\nu, \gamma, \phi,t_{ic})=\frac{q^c_0 h^c(c_i,k_i|\phi,\gamma,t_{ic})+\sum_{j=1}^{n^{*(i)}}n^{*(i)}_j q^c_j\delta_{c^*_j,k^*_j}}{q_0^c+\sum_{j=1}^{n^{*(i)}}n^{*(i)}_j q^c_j}$$
where $q_j^c=1-K_B(t_{ic}|c_j^*,k_j^*)$, and
\begin{eqnarray*}
q_0^c &=& \nu \int_0^\phi \int_0^\infty (1-K(t_{ic}|c,k))G_0(c,k)dcdk\\
&=& \frac{\nu}{\phi \gamma} \int_0^\phi \int_0^\infty\frac{e^{-\frac{k}{\gamma}}}{(1+t_{ic}^c)^k}dkdc\\
&=& \frac{\nu}{\phi \gamma} \int_0^\phi (\frac{1}{\gamma}+ln(1+t_{ic}^{c}))dc
\end{eqnarray*}

Also, the last integration in above equation is computed numerically. By using the property of the censored data we have,
\begin{eqnarray*}
h^c(c_i,k_i|\gamma,\phi, t_{ic})&\propto& (1-K_B(t_{ic}|c_i,k_i))G_0(c_i,k_i)\\
&\propto & [c_i|\gamma,\phi,t_{ic}][k_i|c_i,\gamma,\phi,t_{ic}]\\
&= & \frac{I_{(0,\phi)}(c_i)}{\phi \gamma} \frac{1}{\frac{1}{\gamma}+ln(1+t_{ic}^{c_i})} k_i e^{-k_i(\frac{1}{\frac{1}{\gamma}+ln(1+t_{ic}^{c_i})})}\\
&=&\frac{I_{(0,\phi)}(c_i)}{\phi \gamma} \frac{1}{\frac{1}{\gamma}+ln(1+t_{ic}^{c_i})}\times Gamma(k_i|2,\frac{1}{\frac{1}{\gamma}+ln(1+t_{ic}^{c_i})}).
\end{eqnarray*}
To sample from the first part of the last equation, we use the slice sampling method.
Therefore, by using this MCMC method, we can have a sample from  $h^c(c_i,k_i|\phi,\gamma,t_{ic})$.
Now for observed and censored data, $(c_i,k_i)$ for $i=1,...,n$ can be updated and improved.

In a general form, $(c^*_j,k_j^*)'s$ can be updated on $\phi,\gamma$ and $t$ as the following
\begin{eqnarray}
f(c^*_j,k_j^*| \phi, \gamma, t, n^*)\nonumber
&\propto& G_0(c^*_j,k_j^*|\gamma,\phi)\prod_{\{io:s_{io}=j\}}k_B(t_{io}|c^*_j,k_j^*)\prod_{\{ic:s_{ic}=j\}}(1-K_B(t_{ic}|c^*_j,k_j^*)\nonumber\\
&\propto & [c_j^*|\gamma,\phi,t_{io}][k_j^*|c_j^*,\gamma,\phi,t_{ic}]\prod_{\{ic:s_{ic}=j\} }\frac{1}{(1+t_{ic}^{c_j^*})^{k_j^*}}\nonumber\\
&\propto & {c_j^*}^{n_j^o}I_{(0,\phi)}(c_j^*)\prod_{\{io:s_{io}=j\}}\frac{t_{io}^{c_j^*-1}}{1+t_{io}^{c_j^*}}\times Gamma(n_j^o+1,B^*)
\end{eqnarray}
where $B^*=\sum_{\{io:s_{io}=j\}}(\frac{1}{\gamma}+ln(1+t_{io}^{c_j^*}))+\sum_{\{ic:s_{ic}=j\}}ln(1+t_{ic}^{c_j^*})$ and $n_j^o$ is the number of observed data which located in cluster j.
The important task to generate a sample from equation (9), is drawing from the first part of the equation. Sampling from the gamma distribution is simple.

To sample from
\begin{eqnarray*}
[c_j^*|\phi,\gamma,t]&\propto & {c_j^*}^{n_j^o}I_{(o,\phi)}(c_j^*)\prod_{\{io:s_{io}=j\}}\frac{t_{io}^{c_j^*-1}}{1+t_{io}^{c_j^*}}\\
&\propto & {c_j^*}^{n_j^o}I_{(o,\phi)}c_j^*\prod_{\{io:s_{io}=j\}}(t_{io}^{c_j^*-1})\frac{1}{1+t_{io}^{c_j^*}}
\end{eqnarray*}
it needs to consider auxiliary variables $W=\bigl\{w_{io};\{io:s_{io}=j\}\bigr\}$ such that

$$[c_j^*,W|\phi,t_{io}]={c_j^*}^{n_j^o}I_{(o,\phi)}(c_j^*)\prod_{\{io:s_{io}=j\}} I_{(0, \frac{t_{io}^{c_j^*-1}}{1+t_{io}^{c_j^*}})}(w_{io}).$$
By marginalization over the auxiliary variables, then $[c_j^*|\phi,t_{io}]$ for $j=1,...,n^*$ will be obtained.
Moreover, $w_{io}$'s are uniform variables on $(0,\frac{t_{io}^{c_j^*-1}}{1+t_{io}^{c_j^*}})$.
Therefore we have
$$[c_j^*|\phi,t]={c_j^*}^{n_j^o}I_{(B,\phi)}(c_j^*)$$
where $B=max\{0,\frac{ln(w_{io})}{1+t_{io}}\}$. Now drawing from $[c_j^*|\phi,t]$ is straightforward.

Afterward, using the method that is applied in \cite{escobar95}, $\phi,\gamma$ and $\nu$ will be updated.
If we take $u$ as a latent variable such that
$$[u|\nu,t]= Beta(\nu+1,n)$$
then
$$[\nu|u,n^*,t]=pGamma(a_\nu+n^*,b_\nu-log(u))+(1-p)Gamma(a_\nu+n^*-1,b_\nu-log(u))$$
where $p=\frac{a_\nu+n^*-1}{n(b_\nu-log(u))+a_\nu+n^*-1}$.

And finally to update $\phi$ we have
$$[\phi|c^*,k^*]=[\phi][c^*,k^*|\phi]=\frac{2b_\phi^2}{\phi^3}I_{(b_\phi,\infty)}(\phi)\prod_{j=1}^{n^*}\frac{1}{\phi}I_{(0,\phi)}(c^*)
=\frac{2b_\phi^2}{\phi^{n^*+3}}I_{(b^*,\infty)}(\phi)$$
where $b^*=max\{b_\phi,max_{1\leq j\leq n^*}c_j^*\}$.
So
$$[\phi|c^*,k^*]= Pareto(\phi|2+n^*,b^*).$$
Repeating this technique can update $\gamma$
$$[\gamma|c^*,k^*]=[\gamma]\prod_{j=1}^{n^*}[k_j^*|\gamma]= IGamma(n^*+2,b_\gamma+\sum_{j=1}^{n^*}k_j^*).$$
Now, the all conditional distributions on the equation (8) will be computed.

\section{Data illustrations}
In this section, we consider two kinds of data set, simulated data and real data. For a given sample $(x_1,...,x_n)$, the density function is estimated and compared with the following two density estimators which are used in \cite{Hatjispyros}:

i) The classical kernel density estimation,
$$\tilde{g}_{h}(x;(x_1,...,x_n))\propto n^{-1}\sum_{j=1}^{n}N(x|x_j,h^2)I_{0,+\infty}(x)$$
ii) The kernel density estimation for indirect data,
$$\hat{f}_{J,h}(x;(x_1,...,x_n))\propto n^{-1}\hat{\mu}\sum_{j=1}^{n} x_j^{-1}N(x|x_j,h^2)I_{0,+\infty}(x)$$
where $\hat{\mu}$ is harmonic mean of $(x_1,...,x_n)$.

As we will see, these estimators have good fitness for these type of data and values of $\tilde{g}_{h}$ and $\hat{f}_{J,h}$ are close to the exact values of the true density.
To simulate the required samples, the Gibbs sampler iterates 60,000 times with a burn-in period of 10,000 times.

\subsection{Simulated data}
\subsubsection{Length biased distribution of log-normal}

The first data set simulated from the log-normal distribution with parameter $(\mu,\sigma^2)=(0.5,0.5)$. We use this fact that Length biased distribution of a log-normal with parameters $\mu+\sigma^2$ and $\sigma^2$ is again a log-normal with parameters $\mu$ and $\sigma^2$ \cite{patil78}.

By choosing the log-normal distribution as the kernel, we can show the preference of the model and algorithm. This model tested in \cite{Hatjispyros} for length-biased data using simulated data from the gamma distribution with DPMM when the gamma distribution considered as the kernel.

\begin{figure}[t]
\begin{center}
\includegraphics[width=14cm,height=7cm]{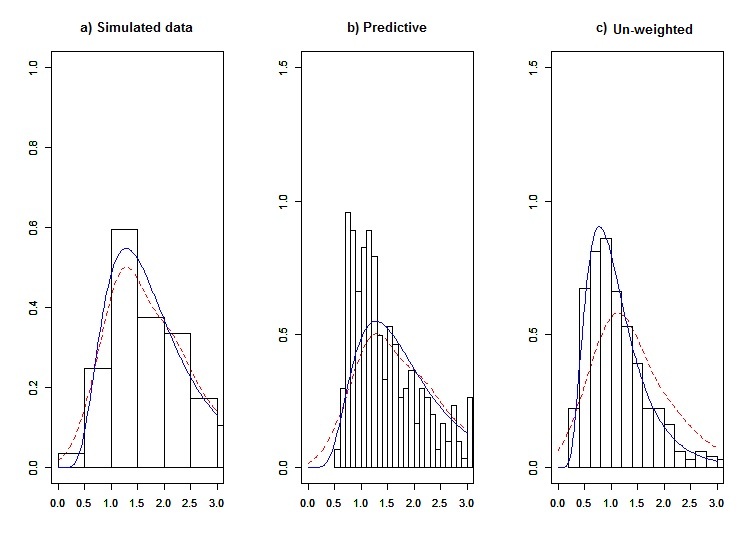}\\
\vspace{-0.2cm}
\caption{\footnotesize{Simulated data from the log-normal distribution with parameters(0.5,0.5) and sample size of n=100. In each figure, the true densities are shown with the solid line and the kernel density estimates $\tilde{g}_{h}$ and $\hat{f}_{J,h}$ with a dashed line.}}
\end{center}
\end{figure}

\begin{figure}[t]
\begin{center}
\includegraphics[width=14cm,height=7cm]{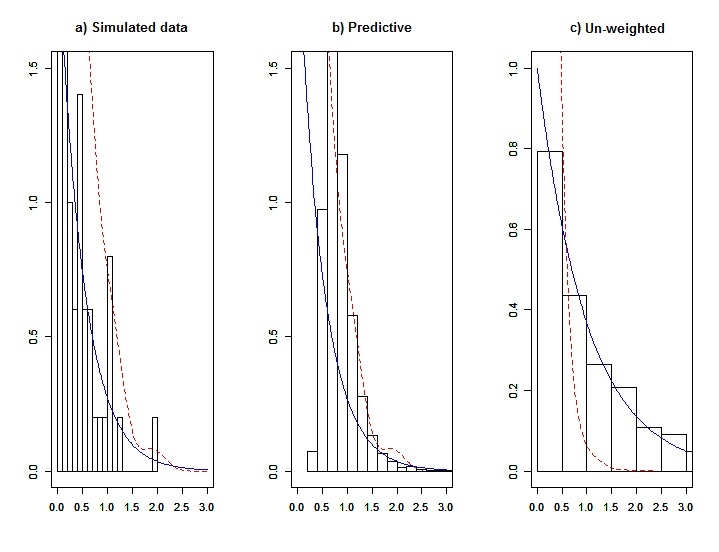}\\
\vspace{-0.2cm}
\caption{\footnotesize{Simulated data from the Weibull distribution with parameter (1,2) and sample size of n=100. In each figure, the true densities are shown with the solid line and the kernel density estimates $\tilde{g}_{h}$ and $\hat{f}_{J,h}$ with a dashed line.}}
\end{center}
\end{figure}

\begin{figure}[t]
\begin{center}
\includegraphics[width=10cm,height=10cm]{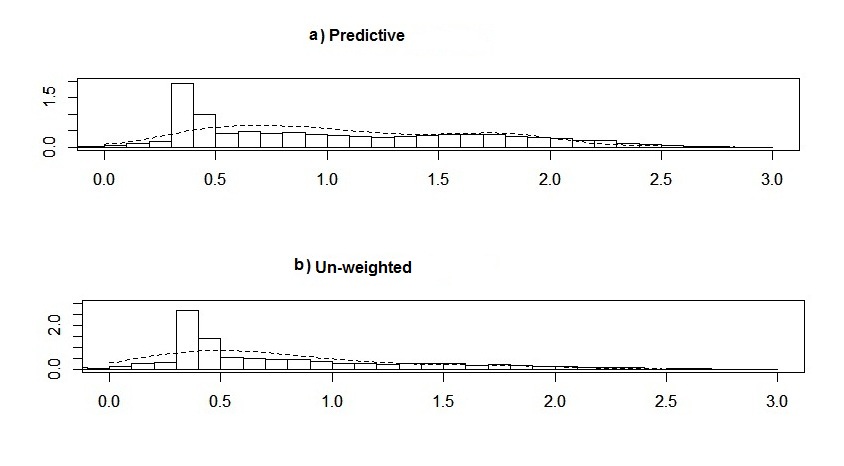}\\
\vspace{-0.2cm}
\caption{\footnotesize{Real data set of the widths of shrubs with size n=46, figure (a) is the histogram of posterior predictive density and $\tilde{g}_{h}$ with the dashed line. Figure (b) is the histogram of de-biased data by using the Metropolis-Hastings algorithm and $\hat{f}_{J,h}$. }}
\end{center}
\end{figure}

\begin{figure}[t]
\begin{center}
\includegraphics[width=10cm,height=5cm]{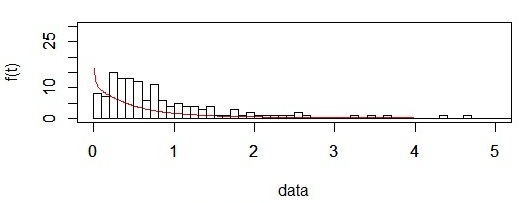}\\
\vspace{-0.2cm}
\caption{\footnotesize{histogram of real data set of bladder cancer patient with size n=137 and estimated curve with the DPBM model.}}
\end{center}
\end{figure}

\begin{figure}[t]
\begin{center}
\includegraphics[width=10cm,height=10cm]{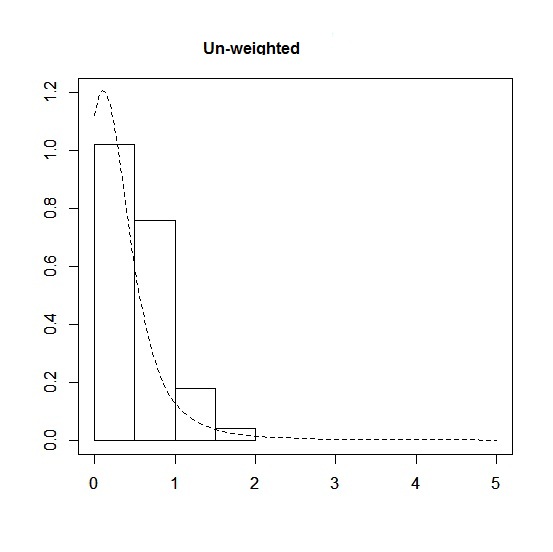}\\
\vspace{-0.2cm}
\caption{\footnotesize{Histogram of un-weighted values of bladder cancer data and $\hat{f}_{J,h}$ curve with the dashed line. }}
\end{center}
\end{figure}

Figure 1 shows this simulated data. In part (a) histogram of simulated data is drawn and also the curve of data is depicted with the solid line and $\tilde{g}_{h}$ is shown with the dashed line.
\newline In part (b) we see the histogram of predictive distribution and also real curve and $\tilde{g}_{h}$ with solid and dashed line respectively.
\newline Part (c) depict the histogram of convert data to the un-weighted version using the Metropolis-Hastings algorithm. Real data here is the log-normal distribution with parameters (0,0.5), that is drawn with the solid line. Kernel density estimator for indirect data is shown with the dash-dot line.
\newline As we see in figure 1, the predictive distribution for simulated data have good fitness and its curve is very similar to the classical kernel density estimator . Distribution of un-weighted data is also close to the real distribution, log-normal distribution with parameters (0,0.5).

\subsubsection{ Weighted distribution of Gamma}
Here we consider a $Gamma(\alpha,\beta)$ distribution with the weight function $w(x|a,b)=x^a exp(\frac{-x}{b})$. So by computing the weighted distribution, we see resulted distribution is again a gamma distribution with parameters $(\alpha +a,\frac{\beta+b}{b\beta})$.
\newline This data set simulated from the $gamma(1,2)$ as the weighted distribution with $w(x)=exp(-x)$(a=0,b=1), and then its un-weighted version is a gamma distribution with (1,1) as the parameters.
\newline In figure 2 part (a), we depict the histogram of the data with its real curve and $\tilde{g}_{h}$.
\newline In part (b), the histogram of predictive values of data, $\tilde{g}_{h}$ and real curve are shown.
\newline In part (c), we can see the histogram of un-weighted distribution that is obtained from the Metropolis-Hastings algorithm and  $\hat{f}_{J,h}$ for this data.

\subsection{Real data}
\subsubsection{Widths of shrubs data}
 For real data, we consider the data that can find in \cite{Muttlak}. This data consists of 46 measurements of widths of shrubs that are sampled by line-transect. In this method of sampling, the probability of inclusion in the sample is proportional to the width of the shrub that it makes it a case of length-biased sampling.
\newline The un-weighted version of values of data and $\hat{f}_{J,h}$ depicted in part (a) of figure 3, also we can see the predictive values of the DPBM model with histogram and $\tilde{g}_{h}$ with the dashed line in part (b).

\subsubsection{Bladder cancer data}
The next real data set is a survival data which included censored values. This data is taken from \cite{lee} which corresponding to remission times (in the month) of a random sample of 138 bladder cancer patients. For fitting the model, we divided data to 10. At first, we apply the DPBM model to the data and in figure 4 draw the estimated density function base on the model and histogram of data.

We consider data that comes from weighted distribution (as in \cite{kilany} a weighted Lomax distribution is fitted to this data set.), then by considering $w(x)=e^{-x}$, the histogram of un-weighted values with the Metropolis-Hastings method and  curve is obtained as the figure 5.

\section{Conclusion}
In this article, we apply the Bayesian nonparametric approach to model weighted data. We use the Dirichlet process mixture model (DPMM) with Burr(XII) distribution as the kernel function in mixing models. We assumed weighted distribution with arbitrary weight function that satisfies in equation(2). By using the Metropolis-Hastings algorithm, the weighted distribution converted to the un-weighted one. As an application, we fit the DPMM with the different kernels and weight functions for real and simulated data sets. As application in the survival study, a real lifetime data set which contained censored observations are used and density and survival functions are calculated.


\end{document}